\documentstyle[11pt,epsfig,epsf]{article}
\def\a{\alpha}  \def\b{\beta}  \def\g{\gamma}  \def\d{\delta}
      \def\ep{\epsilon} 
\def\t{\tau}    \def\n{\nu}    \def\m{\mu}       \def\l{\lambda}
\def\r{\rho}    \def\s{\sigma}    \def\k{\xi}

\def\G{\Gamma}  \def\D{\Delta}

\begin{document}

\title{Effective Two-loop Thermodynamic Potential with Fermions in the 
real-time formalism of thermal field theory
\thanks{Supported by the National Science Foundation of China}}

\author{Wang Xin and Li Jiarong \\  {\normalsize Institute of Particle Physics, 
Huazhong Normal University, Wuhan 430079}}

\date{ }
\maketitle

%%%%%%%%%%%%%%%%%%%%%%%%%%%%%%%%%%%%%%%%%%%%%%%%%%%%%%%%%%%%%%%%%%%%
\vskip-6.5cm
\hskip12cm{\large HZPP-9806}

\hskip12cm{\large March 24, 1998}

\vskip5.5cm
%%%%%%%%%%%%%%%%%%%%%%%%%%%%%%%%%%%%%%%%%%%%%%%%%%%%%%%%%%%%%%%%%%%%
\begin{minipage}{152mm}
\begin{abstract}
Within the real-time formalism (RTF) of thermal field theory, we apply the hard 
thermal loop (HTL) resummation technique to calculating effective two-loop 
thermodynamic potential in quark-gluon plasma (QGP) and its renormalization. 
The result with collective effects is obtained, which is valid for an arbitrary 
number of quark flavors with masses.
\end{abstract}
%\begin{keyword}

\hskip 1cm{\bf Keyword}: quark-gluon plasma (QGP); finite temperature field 
theory

\hskip 1cm{\bf PACS}: 11.10.Wx; 12.38.Mh; 25.75.+r
%\end{keyword}
\end{minipage}

\section{ Introduction}                     
\ \ \ \ \ It is expected that at very high temperature and/or densities 
hadronic matter undergoes the deconfinement transition into a plasma
phase composed of weakly interacting quarks and gluons (QGP). The basic
thermodynamic characteristic of QGP system is its thermodynamic potential, 
and all other thermodynamic quantities of interest, such as the equation 
of state, can be derived from the potential by standard thermodynamic 
relations. In J.I. Kapusta's early works\cite{aaa}, the thermodynamic 
potential of QCD at finite temperature was calculated by using bare 
propagators and vertices. However, without considering the influence from         
collective effects on propagation of partons in QGP, the result is not
suitable for the QGP situation. Recently, P. Arnold and C. Zhai computed 
the free energy density for QCD at high temperature and zero chemical
potential\cite{bbb}, but for the case of massless quarks, and this technique
what they developed seems difficult to deal self-consistently with various 
problems from the collective excitations in hot dense media. 
 
In quark-gluon plasma, deconfined partons are under hot dense condition and
collective effects will affect their dynamic characteristic. For instance, 
interaction of gluons is no longer a long-range force but will change into a
short-range ones. So when calculating thermodynamic potential of the QGP, 
contribution from collective motion should be considered. On the other hand, 
R.D. Pisarski proposed the concept of hard thermal loop (HTL)\cite{lll} and 
from this a systematic method for the calculation of amplitudes in hot gauge 
theories is developed, which is termed the HTL resummation of E. Braaten
-R.D. Pisarski\cite{ccc} and J. Frenkel and J.C. Taylor\cite{ddd}. Because 
the HTL reflects thermal fluctuations fairly well, in this scheme the physical 
effects induced by the collective motion in thermal medium can be discussed 
self-consistently. In this approach, it is natural to distinguish between 
"hard"momenta, $\sim T$, and the "soft" momenta, $\sim gT$. As a consequence, 
for a consistent perturbative calculation, soft momenta lines and vertices 
should be replaced by their effective counterparts in which the HTL is 
introduced. It is well-known that thermodynamic potential is the generating 
functional of vertex function at zero momenta. Obviously, its external lines 
is all soft, hence when performing the calculations of the potential the 
HTL should be taken into account. 

In this paper, we will evaluate the effective two-loop thermodynamic potential 
with $N_{f}$ massive quarks based on the Braaten-Pisarski technique, and we 
give a effective method to deal with the renormalization of this effective 
perturbation theory. We work in the real-time formalism. In order to handle 
the ultraviolet singularities which occur in our procedure, dimensional 
regularization (DR) is used. The $D$ dimensional vector, $P^\m=(p^{0},{\bf p})$, 
is contracted with Minkowski metric, $P^{2}=p_{o}^{2}-p^{2}$ and 
$p=\mid{\bf p}\mid$. For a internal momentum $P$, we introduce the following 
notation:      
$$
  \int (dP)\equiv \int \frac {d^{D}P}{{(2\pi)^{D}}}
$$ 

The plan for the rest of the paper is as follows. Section 2 gives a description 
about HTL gluon self-energy and its effective propagator. Sec.3 computes the
effective thermodynamic potential with fermions in RTF and the renormalization. 
Sec.4 gives a summary and conclusions.

\section { HTL gluon self-energy with massive fermions and gluon effective 
Propagator}   
\ \ \ \ \ Since the momenta of partons in QGP can be soft in our considerations, 
we will need to use effective propagators including the HTL corrections. In 
covariant gauge the effective gluon propagator is decomposed into its 
transverse, longitudinal and gauge components\cite{eee}: 
\begin{eqnarray} 
    -\widetilde G^{\m \n}(K)
                  = A^{\m \n}(K)\D_T(K)
                    +B^{\m \n}(K)\D_L(K)
                    + \k D^{\m \n}(K)\D_G(K) ,  
\end{eqnarray}
where $ \k $ is the gauge parameter, $A^{\m \n}$, $B^{\m \n}$ and 
$ D^{\m \n}$ are the transverse, longitudinal and gauge projectors, and 
\begin{eqnarray}
     \D_{T,L}(K)=\frac {i}{K^2-\Pi _{T,L}(K)}   ,\hskip 1cm
        \D_G(K) =\frac {i}{K^2}     ,
\end{eqnarray}  
where $\Pi_{T,L}(K)$ stand for two independent components, transverse and
longitudinal terms, of the gluon self-energy $\Pi^{\m \n}(K)$.

In the rest-frame of the QGP, using the projectors 
\begin{eqnarray}
     A^{00}(K)&=&A^{i0}(K)=A^{0i}(K)=0,\hskip 1cm 
      A^{ij}(K)=g^{ij}+\frac {k^{i}k^{j}}{k^{2}}, \nonumber \\
     B^{\m \n}(K)&=&g^{\m \n}-A^{\m \n}(K)-D^{\m \n }(K), \hskip 1cm 
      D^{\m \n}(K)={K^{\m }K^{\n}\over K^2} ,
\end{eqnarray}
and the identity
\begin{eqnarray}
    \Pi^{\m \n}(K)=-A^{\m \n}(K)\Pi_T(K)-B^{\m \n}(K)\Pi_L(K),
\end{eqnarray}
we have 
\begin{eqnarray}
 (D-2) \Pi_T (K) &=& -(g^{ij}+ \frac{k^i k^j} {k^2}) \Pi_{ij}(K),\nonumber \\
        \Pi_L(K) &=& -(1- \frac{k_0^2} {k^2}) \Pi_{00}(K), 
\end{eqnarray}
where $ D=4-2\epsilon$ denotes the number of space-time dimensions.
\vskip -110mm
%%%%%%%%%%%%%%%%%%%%%%%%%%%%%%%%%%% self %%%%%%%%%%%%%%%%%%%%%%5
\begin{figure}[h]
\epsfig{figure=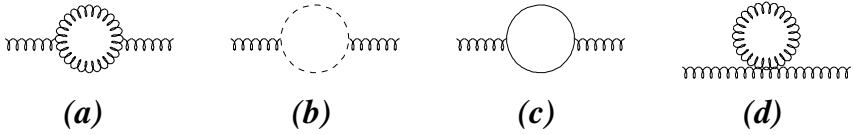,width=15cm}
\vskip -13mm
\caption[]{One-loop gluon self-energy. Curly lines, solid lines, and dashed
lines represent the propagators of gluons, quarks, and ghosts, respectively.}
%\label{fig1}
\end{figure}
%%%%%%%%%%%%%%%%%%%%%%%%%%%%%%%%%%%%%%%%%%%%%%%%%%%%%%%%%%%%%%%%

The gluon self-energy at one-loop level is given in Fig.1. The explicit form
of the bosonic contributions, graph (a), (b),(d) of Fig.1, 
$\Pi^B_{T,L}(k_0,k)$ are\cite{fff}
\begin{eqnarray} 
 \Pi^B_T(k_0,k)&=&\frac{3}{2}m^2_g[x^2+\frac{x(1-x^2)}{2}\ln(\frac{x+1}{x-1})],        
 \nonumber \\
 \Pi^B_L(k_0,k)&=&3m^2_g[(1-x^2)-\frac{x(1-x^2)}{2}\ln(\frac{x+1}{x-1})],
\end{eqnarray}
where $ x\equiv\frac {k_0}{k} $ and $m^2_g\equiv Ng^2T^2/9$ describes the
screening due to thermal pure gluons. 

For the fermionic case, some results obtained come from the massless theory 
only\cite{{lll},{fff}}. Here, we derive the $\Pi^F_{T,L}(k_0,k)$ with fermions 
masses in the HTL approximation. We note that the physical component of the 
fermion propagator at finite temperature has the form
\begin{eqnarray}
S_F(P)&=&(\not\! P+m)[\frac{i}{P^2-m^2}-2\pi\d(P^2-m^2)n_F(P)] \nonumber \\
      &=&(\not\! P+m)(\D^0_F(P)+\D^{\b}_F(P)),
\end{eqnarray}
where $n_F(P)$ represents the fermion distribution function. The temperature
dependent contribution of the fermion loop, graph (c) of Fig.1, to the 
self-energy is given by 
\begin{eqnarray}
    -i\Pi^{(\b)F}_{\m \n}(K)
                 =N_FT_Fg^2\int(dP)
                 {\rm T_r}[\g_{\m}(\not\! P+\not\! K)\g_{\n}\not\! P]
                  [\D_F^0(P+K)\D_F^{\b}+\D_F^{\b}(P+K)\D_F^0(P)]\nonumber\\  
                 =-2i\pi DT_FN_Fg^2\int(dP)
                  \frac{n_F(P)\d(P^2-m^2)}{(P\cdot K)^2-(\frac{K^2}{2})^2}
                  [(P\cdot K)(P_{\m}K_{\n}+P_{\n}K_{\m})-K^2(P_{\m}P_{\n})
                  -g_{\m\n}(P\cdot K)^2]  ,                    
\end{eqnarray}
where the group theory factor $T_F\d^{ab}={\rm T_r}(T^aT^b) $, $N_F$ is a
number of flavors and the terms with two statistical factors has been 
dropped by $\d$-function constraints. From Eq.(5) and (8) one obtains 
$$
     \Pi^F_T(K)
               =\frac {2\pi DT_FN_Fg^2}{D-2}\int(dP)
                \frac {n_F(P)\d(P^2-m^2)}{(P\cdot K)^2-(\frac {K^2}{2})^2}
                [(D-2)(P\cdot K)^2-K^2(p^2-\frac{(\bf {p\cdot k})^2}{k^2})],
$$
\vskip -8mm
\begin{eqnarray}
     \Pi^F_L(K)
               = 2\pi DT_FN_Fg^2\int (dP)\frac {n_F(P)\d(P^2-m^2)(1-x^2)}
                {(P\cdot K)^2-(\frac{K^2}{2})^2}
                [(P\cdot K)^2+K^2p_0^2-2p_0k_0(P\cdot K)].
\end{eqnarray}  

Now we apply the HTL approximation. Let the external momentum $K$ be soft,
$\sim gT$, and the internal ones $P$ be hard, $\sim T$, we can approximate   
\begin{eqnarray}
     \frac {1} {(P \cdot K)^2 -( \frac{K^2}{2})^2 } 
              &\approx& \frac{1}{(P\cdot K)^2} ( 1+ \frac{1}{4} 
                 \frac{(K^2)^2} {(P\cdot K)^2}) \nonumber \\
              &=& \frac {1}{(P\cdot K)^2}(1+O(g^2)) .
\end{eqnarray}

Inserting  Eq.(10) into Eq.(9) and integrating over $p_0$ and $\theta $ 
(${\bf p\cdot k}=pk\cos \theta$), we get the fermionic result with masses: 
\begin{eqnarray}
\Pi_T^F(K)&=&\frac {N_Fg^2}{\pi^2}\int dp pv 
             [x^2+(1-x^2)\frac {x}{2v}\ln(\frac {x+v}{x-v})]n_F(E), \nonumber \\
\Pi_L^F(K)&=&\frac {N_Fg^2}{\pi^2}\int dp pv(1-x^2)
             [1+\frac {x^2-1}{x^2-v^2}-\frac {x}{v}\ln(\frac{x+v}{x-v})]n_F(E) ,
\end{eqnarray}
where $E=\sqrt {m^2+p^2}$, and $v=\frac {p}{E}$ is the velocity of the quarks.
Using Eq.(6) and Eq.(11), one can obtain the HTL gluon self-energy with $N_F$
massive flavors,
\begin{eqnarray}
     \Pi_{T,L}(K)=\Pi_{T,L}^B(K)+\Pi_{T,L}^F(K) .
\end{eqnarray}

Because the poles of $\widetilde G^{\mu\nu}(K)$ correspond to eigenmodes
of the QGP system, Eq.(1) and (12) gives us the desired spectrum of gluon 
collective excitations. For the sake of describing the screening phenomenon 
of thermal gluons, the screening mass is introduced and defined as\cite{iii}
\begin{eqnarray}
          M_{T,L}^2=\lim_{k^2\rightarrow -M_{T,L}^2}\Pi_{T,L}(0,k)
\end{eqnarray}
According to this definition, we have 
\begin{eqnarray}
      \Pi_T(k_0=0,k)
                   &=&0 , \nonumber \\
      \Pi_L(k_0=0,k)
                   &=& M_g^2 \nonumber \\
                   &=&\frac {1}{3}Ng^2T^2+\frac {N_Fg^2}{\pi^2}
                        \int dp(E+\frac {p^2}{E})n_F(E) . 
\end{eqnarray}
It implies static screening of the chromoelectric field but not the 
chromomagnetic field. In order to avoid being in infrared troubles, a magnetic 
mass for the transverse gluon is introduced by assuming that it modifies the 
static behavior in such a way that\cite{ggg} 
\begin{eqnarray}
      \Pi_T(0)= M_m^2 ,
\end{eqnarray}
where $M_m\sim g^2T $.  

Substituting Eq.(12) and (13) into Eq.(2), at finite temperature, one can
obtain the components of the HTL gluon effective propagator:
\begin{eqnarray}
      \D_{T,L}(K)
              &=&\frac {i}{K^2-M^2_{m,g}}+2\pi \d (K^2-M^2_{m,g})n_B(K)
                 \nonumber \\    
              &=&\D_{T,L}^0(K)+\D_{T,L}^{\b }(K), 
                 \nonumber \\
      \D_G(K) &=&\frac {i}{K^2}+2\pi \d (K^2)n_B(K) 
                 \nonumber \\
              &=&\D_G^0(K)+\D_G^{\b}(K) ,
\end{eqnarray}
where $n_B(K)$ represents the boson distribution function.

\section {Effective thermodynamic potential in RTF and its renormalization}
\ \ \ \ \ In this section, we determine the soft exchange corrections to the 
thermodynamic potential in QGP , the relevant diagram to be analyzed is 
depicted in Fig.2, in which the exchanged gluon has momentum components much 
smaller than $T$ while the fermions and other gluons have hard momenta. 
According to the idea of the HTL resummation, for the soft exchanged gluon 
effective propagator $\widetilde G_{\m \n}(K)$ must be used, but the bare ones are 
needed sufficiently for the others, and in our situation no effective vertices 
is needed because the vertices are connected, at least, to one hard line.
\vskip -100mm
%%%%%%%%%%%%%%%%%%%%%%%%%%%%%%%%%%% pot %%%%%%%%%%%%%%%%%%%%%%
\begin{figure}[h]
\epsfig{figure=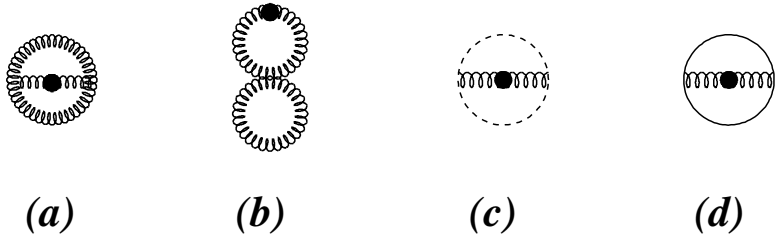,width=15cm}
\vskip -15mm
\caption[]{Effective two-loop thermodynamic potential. The solid blobs indicate the
HTL effective propagator. }
%\label{fig1}
\end{figure}
%%%%%%%%%%%%%%%%%%%%%%%%%%%%%%%%%%%%%%%%%%%%%%%%%%%%%%%%%%%%%%%%
Using the standard Feynman rules of QCD to evaluate the diagram of Fig.2(a), 
one obtains 
$$
       I_a=-C_Ag^2\m ^{4\ep }\int (dK)(dP)
            [-(2K-P)_{\l}g_{\m \n }+(K-2P)_{\n }g_{\l \m }+
            (P+K)_{\m }g_{\n \l }][-g^{\t \n }+(1-\k )D^{\t \n }(K)] 
$$
\vskip -1.3cm
\begin{eqnarray}
         &&\times \D_G(K) [-(2K-P)_{\r }g_{\t \s }+(K-2P)_{\t }
           g_{\s \r }+(P+K)_{\s }g_{\r \t }][-g_{\l \r }+
           (1-\k )D^{\l \r }(P)] \D_G(P) \nonumber \\
         &&\times  [A^{\m \s }(K-P)\D_T(K-P)+B^{\m \s }(K-P)
           \D_L(K-P)  +\k D^{\m \s }(K-P)\D_G (K-P)] \nonumber  \\  
         &=&I_a^{(T)}+I_a^{(L)}+I_a^{(\xi)} ,
\end{eqnarray}
where the group theory factor $C_A\d^{ab}=f^{acd}f^{bcd}$ , $\m $ is a mass 
scale in the DR and  
\begin{eqnarray}
        I_a^{(T)}
               &=&C_Ag^2\m^{4\ep}\int(dK)(dP)
                  [(4D-6)P_{\m}P_{\n}A^{\m\n}(K-P)+(D-2)((2K-P)^2+(2P-K)^2)] 
                  \nonumber \\ 
               &\times& \D_G(K)\D_G(P)\D_T(K-P),
\end{eqnarray}
\vskip -1.1cm
\begin{eqnarray}
        I_a^{(L)}
               &=&C_Ag^2\m^{4\ep}\int (dK)(dP)
                  [(4D-6)P_{\m}P_{\n}B^{\m\n}(K-P)+((2K-P)^2+(2P-K)^2)] 
                  \nonumber \\ 
               &\times& \D_G(K)\D_G(P)\D_L(K-P),
\end{eqnarray}
\vskip -1.2cm
\begin{eqnarray}
      I_a^{(\k)}&=&C_Ag^2\m^{4\ep}\int(dK)(dP)
                   [(4D-6)P_{\m}P_{\n}(g^{\m\n}-D^{\m\n}(K-P))-(D-1)(K+P)^2] 
                   \nonumber \\ 
                &\times& \D_G(K)\D_G(P)\D_G(K-P)
\end{eqnarray}
are the transverse, longitudinal and gauge terms, respectively. 

Substituting Eq.(3), (14) into Eq.(15), one can evaluate these integrals by the 
dimensional regularization method at finite temperature. Since the real part of 
thermodynamic potential is concerned, those terms with three or no statistical 
factors will be dropped\cite{jjj}. In Feynman gauge, including one distribution function term, 
$I_{1\b, a}$, and two distribution functions term, $I_{2\b, a}$, we arrive at
\begin{eqnarray}
      I_{1\b, a}
             =-iC_Ag^2\m {4\ep}\frac {\G (\ep)}{(4\pi)^{2-\ep}}
              \{(\frac {19}{3}+11\ep)(M^2_m)^{1-\ep}\int (dK) \D_T^{\b}(K)
              +(\frac{19}{6}+8\ep)(M_g^2)^{1-\ep}\int (dK)\D_L^{\b}(K)
               \nonumber \\
              +5[2(M_m^2)^{1-\ep}+(M_g^2)^{1-\ep}] \int (dP)\D_G^{\b}(P)
              +10\ep\int (dP)\D_G^{\b}(P)(p^2)^{1-\ep}
               [F(0,M_g^2,p^2)-F(0,M_m^2,p^2)]\},  
\end{eqnarray}
\vskip -1cm     
$$    
    I_{2\b, a}=C_Ag^2\int(dK)(dP)
               \{8M_m^2(2\D_G^0(K)\D_G^{\b}(P)\D_T^{\b}(K-P)+
               \D_T^0(K-P)\D_G^{\b}(P)\D_G^{\b}(K))+i\D_G^{\b}(P) 
$$
\vskip -1cm
\begin{eqnarray}
              &&\times[9(\D_G^{\b}(K)+2\D_L^{\b}(K-P))
               +4(\D_T^{\b}(K-P)-\D_L^{\b}(K-P))
               +7(\D_G^{\b}(K-P)-\D_L^{\b}(K-P))]\nonumber\\
              &&+\frac{3}{2}M_g^2(2\D_G^0(K)\D_G^{\b}(P)\D_L^{\b}(K-P)
               +\D_L^0(K-P)\D_G^{\b}(P)\D_G^{\b}(K))\},
\end{eqnarray}
where function $F(m^2,M_{g,m}^2,p^2)$ is defined as 
$$
    {F(m^2,M_{g,m}^2,p^2)
               \equiv \frac {1}{p^2}\int _0^1dy \frac {(1-y)^{\frac{1}{2}}}{y}} 
               \{[M_{g,m}^2-m^2+\sqrt {(M_{g,m}^2-m^2)^2-4yp^2M_{g,m}^2}] 
$$
\vskip -0.3cm
$$ 
 \times \ln\frac {M_{g,m}^2-m^2-2yp^2+\sqrt {(M_{g,m}^2-m^2)^2-4yp^2M_{g,m}^2}}
 {M_{g,m}^2-m^2+\sqrt {(M_{g,m}^2-m^2)^2-4yp^2M_{g,m}^2}}+[M_{g,m}^2-m^2
$$
\vskip -0.3cm
\begin{eqnarray}
  -\sqrt {(M_{g,m}^2-m^2)^2-4yp^2M_{g,m}^2}]          
  \times \ln\frac {M_{g,m}^2-m^2-2yp^2-\sqrt{(M_{g,m}^2-m^2)^2-4yp^2M_{g,m}^2}}  
   {M_{g,m}^2-m^2-\sqrt {(M_{g,m}^2-m^2)^2-4yM_{g,m}^2p^2}}\}, 
\end{eqnarray}
where $y$ is a Feynman parameter.

It is easy to find that $I_{2\b, a}$ is finite. However $I_{1\b, a}$ contains, 
in addition to finite parts, ultraviolet divergences which take the form of 
multiplying statistic factor. Further more, these overlapping divergences are 
thermal mass dependent, which differs from naive perturbation theory. This 
situation needs a new treatment of the renormalization, and hence we adopt the 
counterterms in Fig.3(a), (b). 
%%%%%%%%%%%%%%%%%%%%%%%%%%%%%%%%%%% sub1 %%%%%%%%%%%%%%%%%%%%%%
\begin{figure}[h]
\epsfig{figure=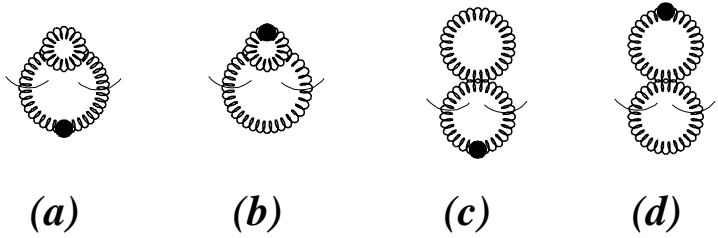,width=15cm}
\vskip -105mm
\caption[]{Renormalization counterterms including three-point gluon vertex and 
four-point gluon vertex}
%\label{fig1}
\end{figure}
%%%%%%%%%%%%%%%%%%%%%%%%%%%%%%%%%%%%%%%%%%%%%%%%%%%%%%%%%%%%%%%%
Where the parentheses indicate that enclosed subgraph is evaluated in the 
vacuum limit, and the particle lines entering and leaving the subgraph are put 
on shell.

Because of using HTL effective perturbation theory, when performing the 
procedure of subtracting divergence, in addition to zero-temperature subgraph 
Fig.3(a), one must consider the contribution from a thermal subgraph, like 
Fig.3(b). And when calculating the vacuum limit of a subgraph, one should save 
the corrections in propagator and vertices from the collective effects induced 
by HTL, and discard the terms with thermal distribution function. 

Considering the Fig.3(a), one can get the pole term of the subgraph easily  
\begin{eqnarray} 
      V_{3g}^0=iC_Ag^2\frac {1}{(4\pi )^2\ep}
                [-\frac {11}{3}P_{\m}P_{\n}+\frac{19}{6}P^2g_{\m\n}].
\end{eqnarray}
For the Fig.3(b), we may multiply by $g_{\m\n}$ to simplify the algebra, the
relevant part of the subgraph is
\vskip -1.1cm
\begin{eqnarray}
      g_{\m\n}I_{3g}^{\b}
              &=&C_Ag^2\m^{4\ep}\int (dQ)\{(4D-6)P_{\l}P_{\r}A^{\l\r}(Q)
                [\D_T^0(Q)-\D_L^0(Q)] \nonumber \\
              &+&[(P+2Q)^2+(Q-P)^2][(D-2)\D_T^0(Q)+\D_L^0(Q)]\}\D_G^0(P+Q).
\end{eqnarray}
Inserting projector $A^{\l\r}(Q)$ into the above equation and with a bit of 
operation, we find that the first term is finite, and applying mass-shell 
condition to momentum $P$, we have the pole term 
\begin{eqnarray}
     g_{\m\n}V_{3g}^{\b}
              =i5C_Ag^2\frac {1}{(4\pi)^2\ep}[2M_m^2+M_g^2].
\end{eqnarray}
When these pole terms are inserted back into the diagram Fig.3, one obtains 
the counterterms of the overlapping divergences of the Fig.2(a)
\begin{eqnarray}
     I_{3g}^{0(div)}
               &=&\int(dP)V_{3g}^0\widetilde G_{\m\n}^{\b}(P) \nonumber \\
               &=&-iC_Ag^2\frac {1}{(4\pi)^2\ep}\int (dP)
               [\frac{19}{3}M_m^2\D_T^{\b}(P)+\frac {19}{6}M_g^2\D_L^{\b}(P)],
\end{eqnarray}
\vskip -1cm
\begin{eqnarray}
    I_{3g}^{\b(div)}
               &=&-\int(dP)V_{3g}^{\b}g_{\m\n}\D_G^{\b}(P) \nonumber \\
               &=&-i5C_Ag^2\frac {1}{(4\pi)^2\ep}
                  [2M_m^2+M_g^2]\int (dP)\D_G^{\b}(P).
\end{eqnarray}
So, after considering the renormalization, total ultraviolet divergences of the 
Fig.2(a) is 
\begin{eqnarray}
     I_a^{(div)}=\frac{1}{2}
                 [I_{1\b, a}^{(div)}-I_{3g}^{0(div)}-I_{3g}^{\b(div)}]
                =0 ,
\end{eqnarray}
finite part is
\begin{eqnarray}
        I_a^{(fin)}=I_{2\b ,a}+I_{1\b, a}^{(fin)},
\end{eqnarray}

The steps to evaluate the Fig.2(b) is similar to the above mentioned, we obtain
\begin{eqnarray}
        I_b&=&-2(D-1)iC_Ag^2\m^{4\ep}\int(dQ)(dP)
              [(D-2)\D_T(P)+\D_L(P)+\D_G(P)]\D_G(Q)\nonumber \\
           &=&I_{1\b, b}+I_{2\b, b},
\end{eqnarray}
where 
\begin{eqnarray}
        I_{1\b,b}=-2iC_Ag^2\m^{4\ep} \frac {\G(-1+\ep)}{(4\pi)^{2-\ep}}
                  [(6-10{\ep})(M_m^2)^{1-\ep}+(3-2{\ep})(M_g^2)^{1-\ep}]
                   \int (dQ)\D_G^{\b}(Q),
\end{eqnarray}
\vskip -1cm
\begin{eqnarray}
        I_{2\b,b}=-2i(D-1)C_Ag^2\m^{4\ep}\int(dQ)(dP)
                  [(D-2)\D_T^{\b}(P)+\D_L^{\b}(P)+\D_G^{\b}(P)]\D_G^{\b}(Q).
\end{eqnarray}

The relevant counterterms are shown in Fig.3(c) and (d). The graph (c) of Fig.3 
gives zero, since its zero-temperature subgraph vanishes in dimensional 
regularization. For Fig.3(d), after straightforward and tedious calculating, 
thermal subgraph yields pole term
\begin{eqnarray}
        V_{4g}^{\b}=-2iC_Ag^2\frac{1}{(4\pi)^2\ep}
                    [g_{\m\n}(2M_m^2+\frac{1}{4}M_g^2)
                     -g_i^{\m}g_j^{\n}g_{ij}\frac{2}{3}(M_m^2-M_g^2)],
\end{eqnarray}
therefore, the subtractive terms are
\begin{eqnarray}
\hskip -7cm   I_{4g}^{0(div)}=0 ,
\end{eqnarray}
\vskip -1cm
\begin{eqnarray}
        I_{4g}^{\b}&=&-\int(dP) V_{4g}^{\b}g_{\m\n}\D_G^{\b}(P) \nonumber \\     
                   &=&2iC_Ag^2\frac{1}{(4\pi)^2\ep}
                      [(6-\frac{8}{3}\ep)M_m^2+(3-\frac{11}{6}\ep)M_g^2],
\end{eqnarray}
then, total overlapping divergences and finite part from Fig.2(b) read 
\begin{eqnarray}
        I_b^{(div)}=\frac{1}{4}
                    [I_{1\b, b}^{(div)}-I_{4g}^{0(div)}-I_{4g}^{\b(div)}]
                   =0      ,
\end{eqnarray}
\vskip -1cm
\begin{eqnarray} 
        I_b^{(fin)}=I_{2\b, b}+I_{1\b, b}^{(fin)}-I_{4g}^{\b (fin)}   .
\end{eqnarray}

As for a ghost loop in Fig.2(c), which its counterterms are shown in Fig.4(a), 
(b), one can get
\begin{eqnarray}
       I_c&=&C_Ag^2\m^{4\ep}\int(dK)(dP)K_{\m}P_{\n}\D_g(K)\D_g(P) \nonumber \\
          && \times [A^{\m\n}\D_T(K-P)+B^{\m\n}(K-P)\D_L(K-P)
             +\D^{\m\n}(K-P)\D_G(K-P)] \nonumber \\
          &=& I_{1\b, c}+I_{2\b, c},
\end{eqnarray}
where $\D_g(K)$ is a ghost propagator, and
\begin{eqnarray}
I_{1\b, c}&=&-iC_Ag^2\m^{4\ep}\frac {\G(\ep)}{(4\pi)^{2-\ep}}\{\frac{1}{6}  
                (1-\ep)(M_m^2)^{1-\ep}\int(dK)\D_T^{\b}(K) \nonumber  \\
          &&+\frac{1}{12}(1+8\ep)(M_g^2)^{1-\ep}\int(dK)\D_L^{\b}(K) \nonumber\\       
          &&+\int(dP)\D_g^{\b}(P)(p^2)^{1-\ep}[F(0,M_m^2,p^2)-F(0,M_g^2,p^2)]\},
\end{eqnarray}
\vskip -1cm
\begin{eqnarray}
 I_{2\b, c}&=& C_Ag^2\m^{4\ep}\int
              (dK)(dP)\{\frac{i}{2}[\D_g^{\b}(P)(\D_G^{\b}(K-P)-\D_L^{\b}(K-P))
              +2\D_g^{\b}(P)\D_L^{\b}(K-P) \nonumber  \\               
           && -\D_g^{\b}(K)\D_g^{\b}(P)] -\frac{M_g^2}{4}[2\D_g^0(K)\D_g^{\b}(P)
              \D_L^{\b}(K-P)+\D_L^0(K-P)\D_g^{\b}(K)\D_g^{\b}(P)] \nonumber  \\
           &&+[p^2-\frac{\bf (p\cdot (k-p))^2}{\bf (k-p)^2}]
             [(\D_L^0(K-P)-\D_T^0(K-P))\D_g^{\b}(K)\D_g^{\b}(P) \nonumber  \\
           &&+2(\D_L^{\b}(K-P)-\D_T^{\b}(K-P))\D_g^0(K)\D_g^{\b}(P)]\}.
\end{eqnarray}
\vskip -95mm
%%%%%%%%%%%%%%%%%%%%%%%%%%%%%%%%%%% sub2  %%%%%%%%%%%%%%%%%%%%%%5
\begin{figure}[h]
\epsfig{figure=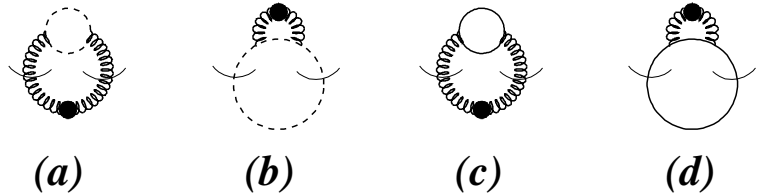,width=15cm}
\vskip -25mm
\caption[]{Renormalization counterterms including ghost-gluon vertex and 
fermion-gluon vertex}
%\label{fig1}
\end{figure}
%%%%%%%%%%%%%%%%%%%%%%%%%%%%%%%%%%%%%%%%%%%%%%%%%%%%%%%%%%%%%%%%
The pole terms and counterterm contributions from Fig.4(a), (b), respectively, 
are 
\begin{eqnarray}
        V_{gG}^0=iC_Ag^2\frac{1}{(4\pi)^2\ep}
                 [\frac{1}{6}K_{\m}K_{\n}+\frac{1}{12}K^2g_{\m\n}], 
\end{eqnarray}
\vskip -1cm
\begin{eqnarray}
        V_{gG}^{\b}=iC_Ag^2\frac {1}{2(4\pi)^2\ep}K^2,
\end{eqnarray}
\vskip -1cm
\begin{eqnarray}
  I_{gG}^{0(div)}&=&\int (dK)V_{gG}^0\widetilde G_{\m\n}^{\b}(K) \nonumber \\
                 &=&iC_Ag^2\frac {1}{(4\pi)^2\ep}\int(dK)
                   [\frac{1}{12}M_g^2\D_L^{\b}(K)+\frac{1}{6}M_m^2\D_T^{\b}(K)],
\end{eqnarray}
\vskip -1cm
\begin{eqnarray}
  I_{gG}^{\b (div)}=\int (dK)V_{gG}^{\b}\D_G^{\b}(K)=0.
\end{eqnarray}
Consequently, we obtain
\begin{eqnarray}
  I_c^{(div)}=\frac{1}{2}I_{1\b, c}^{(div)}-\frac{1}{2}I_{gG}^{0(div)}
               -I_{gG}^{\b (div)}
             =0,
\end{eqnarray}
\vskip -1cm
\begin{eqnarray}
     I_c^{(fin)}=I_{2\b, c}+I_{1\b, c}^{(fin)}.
\end{eqnarray}         

For a soft momentum $(K-P)\sim gT$, we have
\begin{eqnarray}
     n_B(K-P)=\frac {1}{e^{\b\mid{K_0-P_0}\mid}-1}
             \approx \frac {T}{\mid{K_0-P_0}\mid}.
\end{eqnarray}
Substituting Eq.(46) into Eq.(28),(36) and (45) with a little practice, the 
contributions, up to $g^3$, to the thermodynamic potential from bosonic graphs, 
(a),(b) and (c) of Fig.2, together may be written                                                         
\begin{eqnarray}
     I_B(g,M_g(m),T)
               &=&\frac{1}{2}
                   [I_a^{(fin)}+\frac{1}{2}I_b^{(fin)}+I_c^{(fin)}]\nonumber \\
               &=&iC_Ag^2\frac{T^4}{24}
[\frac{1}{24}+\frac{g}{2\pi^2}(9+\frac{11}{2}\frac{M_g}{gT}\arctan\frac{gT}
{M_g})]+O(g^4).
\end{eqnarray}
        
The bosonic sector has been evaluated from the above. Now, we shall incorporate 
fermions into the theory. The diagrams to calculate are as shown in Fig.2(d) 
and Fig.4(c), (d). The loop integrals with fermions can be handled in similar 
manner as bosonic loops. For graph (d) of Fig.2, we write
\begin{eqnarray}
      I_d&=&-T_FN_Fg^2\m^{4\ep}\int(dK)(dP){\rm T_r}[\g_{\m}S_F(K)\g_{\n}S_F(P)]
         \widetilde G^{\m\n}(K-P)\nonumber \\
         &=&I_{1\b, d}+I_{2\b, d},
\end{eqnarray}
here, what remains is the part with one or two statistic factors. Performing 
the trace, applying the Eq.(3) and the dimensional regularization, after a 
lengthy calculation one can get
$$
     I_{1\b,d}=-iT_Fg^2\m^{2\ep}f(4-2\ep)\frac{\G(\ep)}{(4\pi)^2}
               (\frac{4\pi\m^2}{m^2})^{\ep}\{m^2\int(dP)\D_F^{\b}(P) [3+\ep(14 
               -\frac{M_g^2}{m^2}\ln\frac{M_g^2}{m^2} 
$$
\vskip -0.5cm
$$    
               -\frac{M_g}{m^2}\sqrt {M_g^2-4m^2}
                \ln\frac{M_g-\sqrt {M_g^2-4m^2}}{M_g+\sqrt {M_g^2-4m^2}}    
                -(\frac{M_m^2}{m^2}-2)\frac{M_m^2}{m^2}\ln\frac{M_m^2}{m^2}
               -2\frac{M_m^2}{m^2}
$$
$$
                -2\frac{M_m}{m^2}\sqrt {M_m^2-4m^2}
                \ln\frac{M_m-\sqrt {M_m^2-4m^2}}{M_m+\sqrt {M_m^2-4m^2}})] 
                +\frac{M_m^2}{3}\int(dK)\D_T^{\b}(K)
                 [1+\ep(\frac{13}{2}+4\frac{m^2}{M_m^2} 
$$
$$
                +\frac{(2m^2-M_m^2)}{2M_m^3}\sqrt {M_m^2-4m^2}
                \ln\frac{2m^2-M_m^2+M_m\sqrt {M_m^2-4m^2}}
                {2m^2-M_m^2-M_m\sqrt{M_m^2-4m^2}})]+\frac{M_g^2}{6}
                \int(dK)\D_L^{\b}(K)
$$
$$
                \times [1+\ep(\frac{5}{3}-8\frac{m^2}{M_g^2}  
                +\frac{1}{2M_g}\sqrt {M_g^2-4m^2}
                \ln\frac{2m^2-M_g^2+M_g\sqrt
                 {M_g^2-4m^2}}{2m^2-M_g^2-M_g\sqrt{M_g^2-4m^2}})] 
$$
\vskip -0.8cm
\begin{eqnarray}
                +\frac{2}{3}\ep\int (dP)\D_F^{\b}(P)p^2
                [F(m^2,M_g^2,p^2)-F(m^2,M_m^2,p^2)]\} ,
\end{eqnarray}
\vskip -1cm
$$
    I_{2\b, d}=T_Fg^2\m^{4\ep}f(D)\int(dK)(dP)\{i(D-2)
      [\D_F^{\b}(K)\D_F^{\b}(P)-(\D_T^{\b}(K-P)+\D_L^{\b}(K-P))\D_F^{\b}(P)]
$$
$$
      +(2m^2+\frac{D-2}{2}M_g^2)(\D_L^0(K-P)\D_F^{\b}(K)\D_F^{\b}(P)
      +2\D_F^0(K)\D_L^{\b}(K-P))\D_F^{\b}(P))
$$
$$
      +\frac{D-2}{2}M_m^2(\D_T^0(K-P)\D_F^{\b}(K)\D_F^{\b}(P)
      +2\D_F^0(K)\D_T^{\b}(K-P)\D_F^{\b}(P))
$$ 
$$
      +2[p^2-\frac{\bf (p\cdot (k-p))^2}{\bf (k-p)^2}]
        [(\D_L^0(K-P)-\D_T^0(K-P))\D_F^{\b}(K)\D_F^{\b}(P)
$$
\vskip -0.8cm
\begin{eqnarray}
     +2(\D_L^{\b}(K-P)-\D_T^{\b}(K-P))\D_F^0(K)\D_F^{\b}(P)]\} .
\end{eqnarray}
The pole terms and counterterm contributions from Fig.4(c), (d),
respectively, are
\begin{eqnarray}
     V_{FG}^0=-iT_Fg^2\frac{1}{(4\pi)^2\ep}(-\frac{4}{3}K_{\m}K_{\n}
              +\frac{4}{3}K^2g_{\m\n}) ,
\end{eqnarray}
\vskip -0.8cm
\begin{eqnarray}
     V_{FG}^{\b}=-i(T^aT^a)_{\a\b}g^2\frac{1}{(4\pi)^2\ep}(\not \!P-4m) ,
\end{eqnarray}
\vskip -0.8cm
\begin{eqnarray}
     I_{FG}^{0(div)}&=&\int(dK)V_{FG}^0\widetilde G_{\m\n}^{\b}(K) \nonumber\\
                    &=&-iT_Fg^2\frac{1}{(4\pi)^2\ep}\frac{4}{3}\int(dK)
                       [2M_m^2\D_T^{\b}(K)+M_g^2\D_L^{\b}(K)] ,
\end{eqnarray}
\vskip -0.8cm
\begin{eqnarray}
    I_{FG}^{\b(div)}&=&-\int(dP){\rm T_r}[V_{FG}^{\b}(P)(\not \! P+m)]\D_F^{\b}(P) 
                          \nonumber\\
                    &=&-i3m^2T_Fg^2\frac{1}{(4\pi)^2\ep}f(4-2\ep)
                          \int(dP)\D_F^{\b}(P) .
\end{eqnarray}

As a result total ultraviolet overlapping divergences of the Fig.2(d), with 
considering the renormalization, is
\begin{eqnarray}
     I_d^{(div)}=\frac{1}{2}I_{1\b, d}^{(div)}-\frac{1}{2}I_{FG}^{0(div)}
                 -I_{FG}^{\b(div)}
                =0 .
\end{eqnarray}
To compute the contribution from finite parts, we take the limit
$\ep\rightarrow 0$. Integrating over $p_0$ and $k_0$ and changing the variable 
of integration by letting $\bf{(k-p)\rightarrow k}$, we get
$$
      I_{2\b, d}=2iT_Fg^2\{2\int_{p,k(hard)}(dp)(dk)\frac{n_F(E_p)}{E_p}
                 \frac{n_F(E_k)}{E_k}+\int_{p(hard)}(dp)\int_{k(soft)}(dk)\{
                 \frac{n_F(E_p)}{E_p}[\frac{n_F(E_{p+k})}{E_{p+k}} 
$$
$$
                \times [(2m^2+M_g^2+2(p^2-\frac{\bf( p\cdot k)^2}{k^2}))
                (\frac{1}{(E_p+E_{p+k})^2-k^2-M_g^2}+\frac{1}{(E_p-E_{p+k})^2
                -k^2-M_g^2})
$$
\vskip -0.5cm
\begin{eqnarray}
                +(M_m^2-2(p^2-\frac{\bf( p\cdot k)^2}{k^2}))(\frac{1}
                {(E_p+E_{p+k})^2-k^2-M_m^2}+\frac{1}{(E_p-E_{p+k})^2-k^2-M_m^2})]] 
                -[\frac{n_B(E_k^{(L)})}{E_k^{(L)}}    
\end{eqnarray}
$$
                \times (2m^2+M_g^2+2(p^2-\frac{\bf( p\cdot k)^2}{k^2}))
                (\frac{1}{(E_p+E_k^{(L)})^2-({\bf {k+p})^2}-m^2}
                +\frac{1}{(E_p-E_k^{(L)})^2-({\bf {k+p})^2}-m^2}) 
$$
$$
                +\frac{n_B(E_k^{(T)})}{E_k^{(T)}}
                (M_m^2-2(p^2-\frac{\bf(p\cdot k)^2}{k^2}))
                (\frac{1}{(E_p+E_k^{(T)})^2-({\bf {k+p})^2}-m^2}
                +\frac{1}{(E_p-E_k^{(T)})^2-({\bf {k+p})^2}-m^2})]\}\},
$$
where 
$$ 
        E_k^{(T,L)}=\sqrt {k^2+M_{m,g}^2}, \hskip  1cm  
        \int (dk)\equiv \int\frac {d^3k}{(2\pi)^3}.
$$ 
When momentum $k$ is soft, we can approximate
$$
        E_{k+p}+E_p\approx 2E_p+\frac{\bf k\cdot p}{E_p},\hskip 1cm 
        E_{k+p}-E_p\approx \frac{\bf k\cdot p}{E_p},
$$ 
\vskip -0.8cm
\begin{eqnarray}
        n_F(E_{k+p})=\frac {1}{e^{\b E_{k+p}}+1}\approx n_F(E_p).
\end{eqnarray}

Substituting Eq.(57), (46) into Eq.(56), after a tediously long calculation we 
obtain, up to $g^3$, contributions from the loop with massive fermions 
\begin{eqnarray} 
  I_F(g,m,M_g(m),T)&=&\frac {1}{2}[I_{2\b, d}+I_{1\b, d}^{(fin)}] \nonumber\\ 
                   &=&iT_FN_F\{2g^2I_1^2(m,\b)-\frac{2T}{(2\pi)^2}g^3[2m^2(1
                     -\frac{M_g}{gT}{\arctan\frac{gT}{M_g}})I_2(m,\b) \nonumber\\
                   &&-(1+\frac{1}{3}(\frac{gT}{M_g})^2+\frac{M_g}{gT}
                     {\arctan\frac{gT}{M_g}})I_3(m,\b)+\frac{2}{3}
                    (\frac{M_g}{gT}+\frac{1}{2}\frac{gT}{M_g})I_4(m,\b)] \\
                   &&+\frac{T^2}{(2\pi)^2}g^3[m^2((\frac{gT}{M_g})^2
                     -\frac{3}{2}(1-\frac{M_g}{gT}{\arctan\frac{gT}{M_g}}))
                      I_5(m,\b)+I_1(m,\b)]\}+O(g^4),\nonumber
\end{eqnarray}
where 
$$
      I_1(m,\b)=\frac{2}{(2\pi)^2}\int_0^{\infty}dp
                \frac{p^2}{\sqrt {m^2+p^2}}\frac {1}{e^{\b \sqrt {m^2+p^2}}+1},
$$
$$
      I_2(m,\b)=\frac{2}{(2\pi)^2}\int_0^{\infty}dp 
                \frac{p^2}{m^2+p^2}\frac {1}{(e^{\b \sqrt {m^2+p^2}}+1)^2},
$$
$$
      I_3(m,\b)=\frac{2}{(2\pi)^2}\int_0^{\infty}dp
                \frac{p^4}{m^2+p^2}\frac {1}{(e^{\b \sqrt {m^2+p^2}}+1)^2},
$$
$$
      I_4(m,\b)=\frac{2}{(2\pi)^2}\int_0^{\infty}dp
              \frac{p^3}{\sqrt{m^2+p^2}}\frac {1}{(e^{\b \sqrt {m^2+p^2}}+1)^2},
$$
\vskip -0.5cm
\begin{eqnarray}
     I_5(m,\b)=\frac{2}{(2\pi)^2}\int_0^{\infty}dp
           \frac{p^2}{(m^2+p^2)^\frac{3}{2}}\frac {1}{e^{\b \sqrt {m^2+p^2}}+1}.
\end{eqnarray}
Taking some proper approximations, such as a high-temperature expansion, these 
integrals in Eq.(59) can be evaluated.

Combining Eq.(58) and (47), finally we get the effective two-loop thermodynamic potential 
in QGP
\begin{eqnarray}
        \Omega_{exch}=-i[I_F(g,m,M_g(m),T)+I_B(g,M_g(m),T)],
\end{eqnarray}
in which the effects both of finite fermion masses and of collective motion
induced by the HTL are involved. Using this expression for the thermodynamic 
potential, we can consider the limiting case of our interest.

In the ultra-relativistic $(m=0)$ limit, Eq.(60) yields  
\begin{eqnarray}
  \Omega_{exch}^{ultr}
            &=&g^2(\frac {T^2}{24})^2(C_A+2T_FN_F)+g^3(\frac {T^2}
               {2\pi})^2\{[9+\frac {11}{2}\sqrt {\frac {1}{3}(N+\frac {N_F}{2})}
               {\arctan\sqrt {\frac {6}{2N+N_F}}}]\frac {C_A}{12}\nonumber \\
            &&+[\frac{1}{2}+\frac{1}{\pi^2}(1+\frac {2}{2N+N_F}-\frac {2}{3}
               \sqrt {\frac {6}{2N+N_F}}+\sqrt {\frac {1}{3}(N+\frac {N_F}{2})}
              (\arctan\sqrt {\frac {6}{2N+N_F}}-\frac {2}{3}))\nonumber \\
            &&\times(\frac {\pi^2}{6}+\frac {3}{2}\zeta(3)-2)]T_FN_F\}+O(g^4).
\end{eqnarray}
This result agrees with the one found in the imaginary time formalism (ITF)
\cite{bbb}, which is to be expected because of our considering only HTL in the 
static limit.

\section {Summary and conclusions}
\ \ \ \ \ In the present paper, we have derived the HTL gluon self-energy with 
fermion masses and its effective propagator by use of the RTF. we have 
calculated explicitly effective two-loop thermodynamic potential in QGP, and 
have made a detailed presentation of its renormalization. we obtain the result
including collective effects due to the HTL, which is suitable for an arbitrary 
number of quark flavors with masses. 

In particular, we have found that the effective two-loop thermodynamic 
potential in QGP is composed of two parts. One is the contribution, $g^2$ 
terms, from interactions between the hard bare partons, which is just the 
result in naive perturbation theories. The other part is the higher order 
terms, which is owing to the interactions between soft gluon and hard partons 
in hot dense media, and chromomagnetic mass effects is absent from the $g^3$ 
level. 

To be sure, after introducing the THL, one can still apply the DR to handle 
self-consistently the regularization and renormalization of the thermodynamic 
potential, but a point which is especially to be noted is that, within this 
effective perturbation theory scheme, we should introduce the thermal 
counterterms and employed the RTF, which makes our calculations simple and 
clear. 

Finally, it will be straightforward to evaluate the thermodynamic potential to 
higher order making use of the methods and formulas developed in this paper.

\end{document}